\documentclass[12pt]{iopart}
\usepackage{citesort}
\usepackage{epsfig}
\usepackage{iopams}
\usepackage{tabulary}
\usepackage[table]{xcolor}
\definecolor{Gray}{gray}{0.85}

\begin{document}

\title[A cCPT optomechanical scheme for accessing the ultra strong coupling regime] {A cavity-Cooper pair transistor scheme for investigating quantum optomechanics in the ultra-strong coupling regime}

\author{A.J. Rimberg$^{1}$, M.P. Blencowe$^{1}$, A.D. Armour$^{2}$, and P.D. Nation$^{3}$}
\address{${}^{1}$Department of Physics and Astronomy, 6127 Wilder Laboratory,
Dartmouth College, Hanover, NH 03755, USA\\
${}^{2}$School of Physics and Astronomy, University of Nottingham,
Nottingham,\\ NG7 2RD, UK\\
${}^{3}$Department of Physics, Korea University, Seoul 136-713, Korea}
\ead{\mailto{alexander.j.rimberg@dartmouth.edu}}

\begin{abstract}
We propose a scheme involving a  Cooper pair transistor (CPT) embedded in a superconducting microwave cavity, where the CPT serves as a
charge tunable quantum inductor to facilitate ultra-strong  coupling between photons in the cavity and a nano- to meso-scale mechanical resonator.
The mechanical resonator is capacitively coupled to the CPT, such that mechanical displacements of the resonator
cause a shift in the CPT inductance and hence the cavity's resonant frequency. The  amplification provided by the CPT is sufficient for
 the zero point motion of the mechanical resonator alone to cause a significant change in the cavity resonance. Conversely,
a single photon in the cavity causes a shift in the mechanical resonator position on the
order of its zero point motion. As a result, the cavity-Cooper pair transistor (cCPT) coupled to a mechanical resonator
will be able to access a regime in which single photons can affect single phonons and vice
versa. Realizing this ultra-strong coupling regime will facilitate the
creation of non-classical states of the mechanical resonator, as well as the means to accurately
characterize such states by measuring the cavity photon field.
\end{abstract}

\submitto{NJP} \maketitle

\section{Introduction}

There is presently intense worldwide interest in the application of quantum mechanical phenomena to communications, information processing, and precision measurement.  At the level of atoms, photons and even molecules the laws of quantum mechanics clearly hold sway.  In contrast, macroscopic objects are just as clearly described by Newtonian mechanics.  Practitioners of the above fields are therefore keenly interested in the boundary between quantum mechanical and classical behavior, and in the ways in which quantum behavior can be extended into regimes that at first glance might seem to lie in the province of classical mechanics~\cite{Haroche:2006,Aspelmeyer:2013}.

One field centered around the connection between the quantum and classical worlds is that of cavity optomechanics~\cite{Aspelmeyer:2013,Poot:2012}.  Motivated by a desire to observe and control quantum phenomena in mechanical structures, many researchers have focussed on the idea of coupling a mechanical resonator to an optical or microwave cavity.  If motion of the mechanical resonator shifts the cavity's resonant frequency (by changing the cavity length, for instance), then phase sensitive optical measurements of the cavity can be used to measure the resonator position.  There has been a wealth of recent results in this area~\cite{Teufel:2011a,Chan:2011,Purdy:2013,SafaviNaeini:2013,Murch:2008,Brennecke:2008,Bochmann:2013,Palomaki:2013}, including cooling mechanical resonators to their quantum ground state~\cite{Teufel:2011a,Chan:2011}, observation of radiation pressure shot noise~\cite{Purdy:2013}, production of squeezed light by a mechanical resonator~\cite{SafaviNaeini:2013} and  the optomechanics of cold atoms~\cite{Murch:2008,Brennecke:2008}.

The quantum dynamics of cavity optomechanical systems are usually described by the Hamiltonian
\begin{equation}
{\cal{H}}_{\mathrm{OM}} = \hbar\omega_0 a^{\dagger}a  + \hbar\omega_{m} b^{\dagger}b + \hbar g_0 a^{\dagger}a(b+b^{\dagger}),
\label{homeq}
\end{equation}
where $\omega_0$ is the cavity mode frequency, $\omega_{m}$ is the frequency of the mechanical resonator, $a$ and $a^{\dagger}$ are the cavity photon annihilation and  creation operators, and $b$ and $b^{\dagger}$ are the associated phonon annihilation and creation operators.  The first two terms of ${\cal{H}}_{\mathrm{OM}}$ describe harmonic motion of the cavity and mechanical resonator, while the last term describes a dispersive shift in the cavity frequency due to mechanical motion.  The parameter $g_{0}$ is the vacuum optomechanical coupling strength, and expresses the shift in cavity frequency due to displacement of the mechanical resonator by its zero point length $x_{\mathrm{zp}}=\sqrt{\hbar/2m\omega_{m}}$.  Essentially, $g_{0}$ describes the strength of interaction between a single photon and a single phonon.

An exciting experimental challenge facing the cavity optomechanics community is reaching the ultra-strong optomechanical quantum regime, for which the coupling term in ${\cal{H}}_{\mathrm{OM}}$ becomes important at the scale of individual quanta~\cite{Ludwig:2008,Nunnenkamp:2011,Nation:2013,Lorch:2013}.
There are two main requirements to reach this regime.  First, the shift in cavity frequency due to a single phonon must be larger than the linewidth $\kappa = \omega_0/Q$, where $Q$ is the cavity mode quality factor; this is equivalent to requiring that the ratio $g_0/\kappa$, called the granularity parameter, be greater than one~\cite{Murch:2008}.  Second,  the displacement of the mechanical resonator due to the force of a single photon must be greater than the zero point displacement  $x_{\mathrm{zp}}$; equivalently, the ratio $2g_0/\omega_m$ must also be greater than one~\cite{Nunnenkamp:2011,Aspelmeyer:2013}.   In terms of a single parameter, it is convenient to consider the product $g_0^{2}/(\kappa\omega_m)$; if this parameter is greater than one, then we are in the  single-photon strong-coupling regime~\cite{Nation:2013,Lorch:2013}.   In table~\ref{omtab} we show a range of values for these parameters that have been realized in recent optomechanics experiments.
 \begin{table}
 \begin{center}
 \begin{tabulary}{\linewidth}{LCCCCC}
 \rule[-2mm]{0mm}{7mm} {\bfseries System} & $N$ & $\frac{\omega_m}{\kappa}$ & $\frac{g_{0}}{\kappa}$ &$\frac{g_{0}}{\omega_m}$ & $\frac{g_{0}^{2}}{\kappa\omega_m}$ \\
 \hline
\rule[-2mm]{0mm}{6mm}Superconducting $LC$ oscillator~\cite{Teufel:2011a} & $1\e{11}$ & 60 & $3\e{-3}$ & $4\e{-5}$ & $1\e{-7}$ \\
\hline
\rule[-2mm]{0mm}{6mm}Si optomechanical crystal~\cite{Chan:2011} & $6\e{9}$ & 7 & $2\e{-3}$ & $2.5\e{-4}$ & $5\e{-7}$ \\
\hline
\rule[-2mm]{0mm}{6mm}Cold atomic gas~\cite{Murch:2008} & $4\e{4}$ & 0.06 & 22 & 340 & 7,500\\
\hline
\rowcolor{Gray}\rule[-2mm]{0mm}{6mm}cCPT-mechanical resonator & $5\e{9}$ & 20 & 8 & 0.4 & \textcolor{red}{\bfseries 3}  \\
 \hline
\end{tabulary}
\end{center}
 \caption{\label{omtab} The sideband ratio $\omega_m/\kappa$, the granularity parameter $g_0/\kappa$, the backaction parameter $g_0/\omega_m$ and the combined quantum nonlinearity parameter $g_0^{2}/\kappa\omega_m$, for certain demonstrated opto- and electromechanical systems, where $N$ is the estimated number of atoms making up the mechanical resonator.  Also shown for comparison are the estimated parameters of the cCPT-mechanical resonator scheme discussed in the present work.}
\end{table}

In the present work, we describe an optomechanical scheme involving a Cooper pair transistor (CPT) that is embedded in a superconducting microwave cavity,
where a mechanically compliant, biased gate electrode couples mechanical motion to the cavity via the CPT. The basic scheme for the cavity-CPT-mechanical resonator (cCPT-MR) system is given in figures~\ref{mrfig} and~\ref{rmfig}. In particular, we will show that the cCPT-MR device is capable of attaining the ultra-strong coupling regime, with relevant achievable  parameters given in table~\ref{omtab}.  Note that reference~\cite{heikkila:2013} discusses a very similar scheme. There was also an earlier proposal to enhance effective optomechanical coupling strengths in the microwave regime by mediating the coupling through a SQUID~\cite{Blencowe:2007}.

This paper is organized as follows. In section \ref{Sec:2} we
describe the cCPT-MR device and give a physical derivation of the effective optomechanical coupling strength  $g_0$ of the device. Next in section \ref{Sec:derivation} we give a more systematic derivation of the optomechanical Hamiltonian~(\ref{homeq}), starting with a circuit model of the cCPT-MR device. Finally, in  section \ref{Sec:disc}, we conclude with a discussion of our results and future work. The appendix contains the derivation of the circuit model.

\section{The cCPT-MR Device}
\label{Sec:2}
\begin{figure}
\begin{center}
\includegraphics[width=8.5cm]{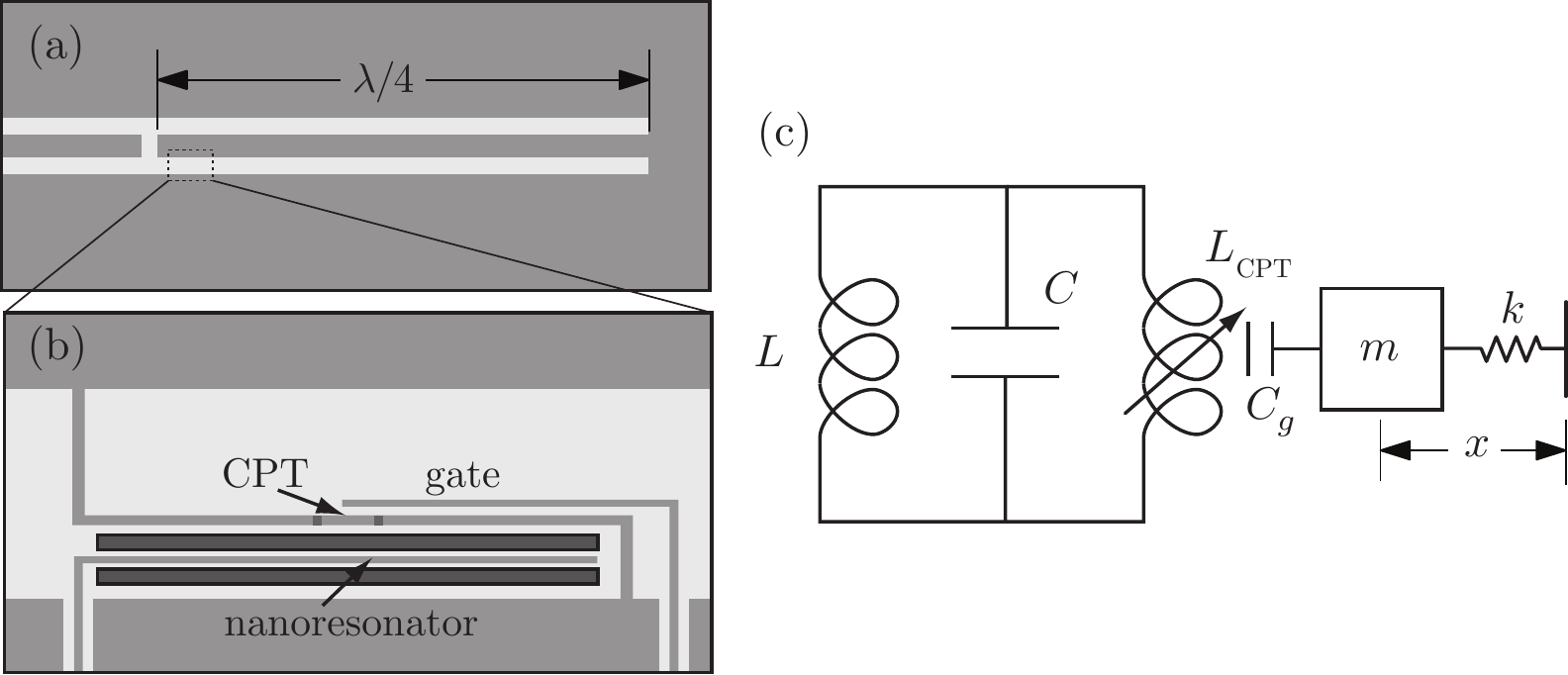}
\caption{\label{mrfig} (a) Schematic illustration of a shorted $\lambda/4$, microwave resonator coupled to a feedline. (b) Detail of the CPT location and the method of coupling to a mechanical resonator. (c) Simplified circuit diagram of the  device. }
\end{center}
\end{figure}

\begin{figure}
\begin{center}
\includegraphics[width=7.5cm]{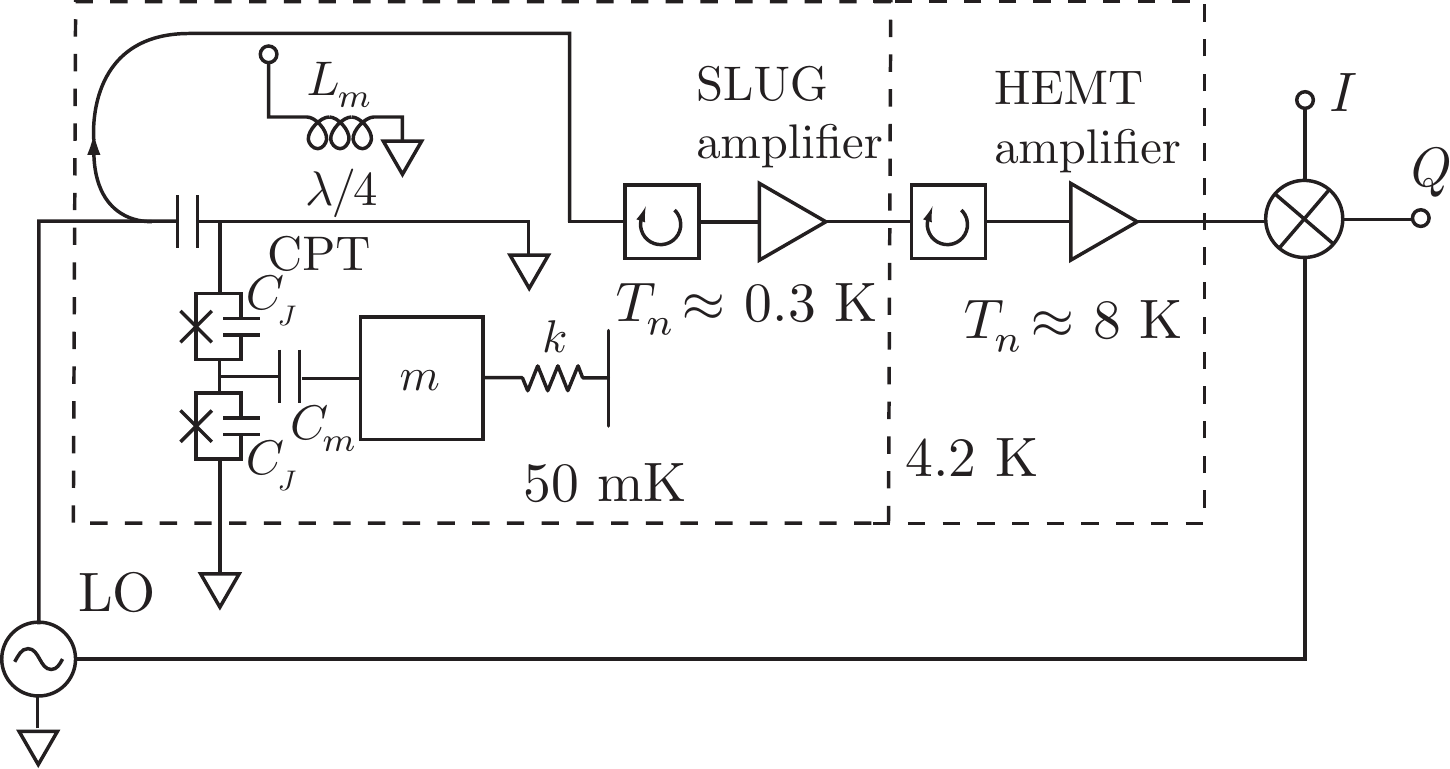}
\caption{\label{rmfig} Schematic diagram of a cCPT coupled to a mechanical resonator, with amplifier stages for  measuring the cavity photon field. A flux bias is applied to the cCPT to control the total CPT phase; a separate gate line (not shown) controls the island charge. }
\end{center}
\end{figure}

Referring to figures~\ref{mrfig} and~\ref{rmfig}, the cCPT comprises two discrete components.  One, the Cooper pair transistor (CPT), consists of a small superconducting island in the Coulomb blockade regime that is coupled via two Josephson junctions to macroscopic superconducting leads.  The CPT has been extensively studied~\cite{Tuominen:1992,Matveev:1993,Eiles:1994,Joyez:1994,Joyez:1995}, and its properties are now well understood. The second component of the cCPT is a shorted quarter-wave, superconducting high-$Q$ microwave cavity, which is flux biased to allow control over the total dc cCPT phase.   The  microwave cavity, made from a transmission line of impedance $Z_{0}$, is based on the circuit QED architecture~\cite{Wallraff:2004,Blais:2004} that has led to significant advances in the coherence and control of quantum superconducting circuits.  The cCPT is created by embedding the CPT at the open end of the center conductor (a voltage antinode), so that it connects the central conductor of the cavity to the ground plane.

For our purposes, the CPT is well described by considering two charge states, $|0\rangle$ and $|1\rangle$, corresponding to zero and one excess Cooper pairs on the island.  These charge states are separated by an electrostatic energy difference $2\varepsilon=4E_c(1-n_g)$ dependent on gate charge $n_g$, and are coupled to each other via the Josephson energy $E_J$. Introducing cavity photon annihilation and  creation operators $a$ and $a^{\dagger}$, the Hamiltonian of the cCPT can be expressed as (see appendix):
\begin{equation}\label{heq}
{\mathcal{H}}_{\mathrm{cCPT}} = \hbar\omega_0 a^{\dagger}a + \varepsilon \sigma_{z} - E_J \sigma_{x} \cos\left[\Delta_0 (a + a^{\dagger}) +\pi\Phi_{\mathrm{ext}}/\Phi_0\right],
\end{equation}
where $\sigma_{x}$ and $\sigma_{z}$ are the Pauli matrices, $\omega_0$ is the cavity frequency, $\Phi_{\mathrm{ext}}$ is an external flux bias, and $\Phi_{0}$ is the flux quantum. The first two terms in equation~(\ref{heq}) describe the cavity photons and the CPT charge.  The third term describes the coupling between the CPT charge states and the cavity photons. In a standard CPT, this term would read $E_J \sigma_{x}\cos\varphi/2$ where $\varphi$, the total superconducting phase difference between the source and drain, can be treated as a classical variable~\cite{Joyez:1994,Joyez:1995}. In the cCPT, however, quantum fluctuations of the cavity photon field must be accounted for via the identification $\hat{\varphi} /2= \Delta_0 (a+a^{\dagger})$, which is proportional to the electric field in the cavity at the location of the CPT.  The dimensionless parameter $\Delta_0=\sqrt{Z_0/R_K}\ll 1$, where $R_K=h/e^{2}={25.8}~{\mathrm{k}}\Omega$ is the resistance quantum, describes the strength of the quantum phase fluctuations of the cavity field, which can be important for large cavity photon numbers~\cite{Blencowe:2012,Chen:2013}.  Experimental study~\cite{Chen:2013,Chen:2013a} indicates that equation~(\ref{heq}) accurately models the cCPT.

\begin{figure}
\begin{center}
\includegraphics[width=8.5cm]{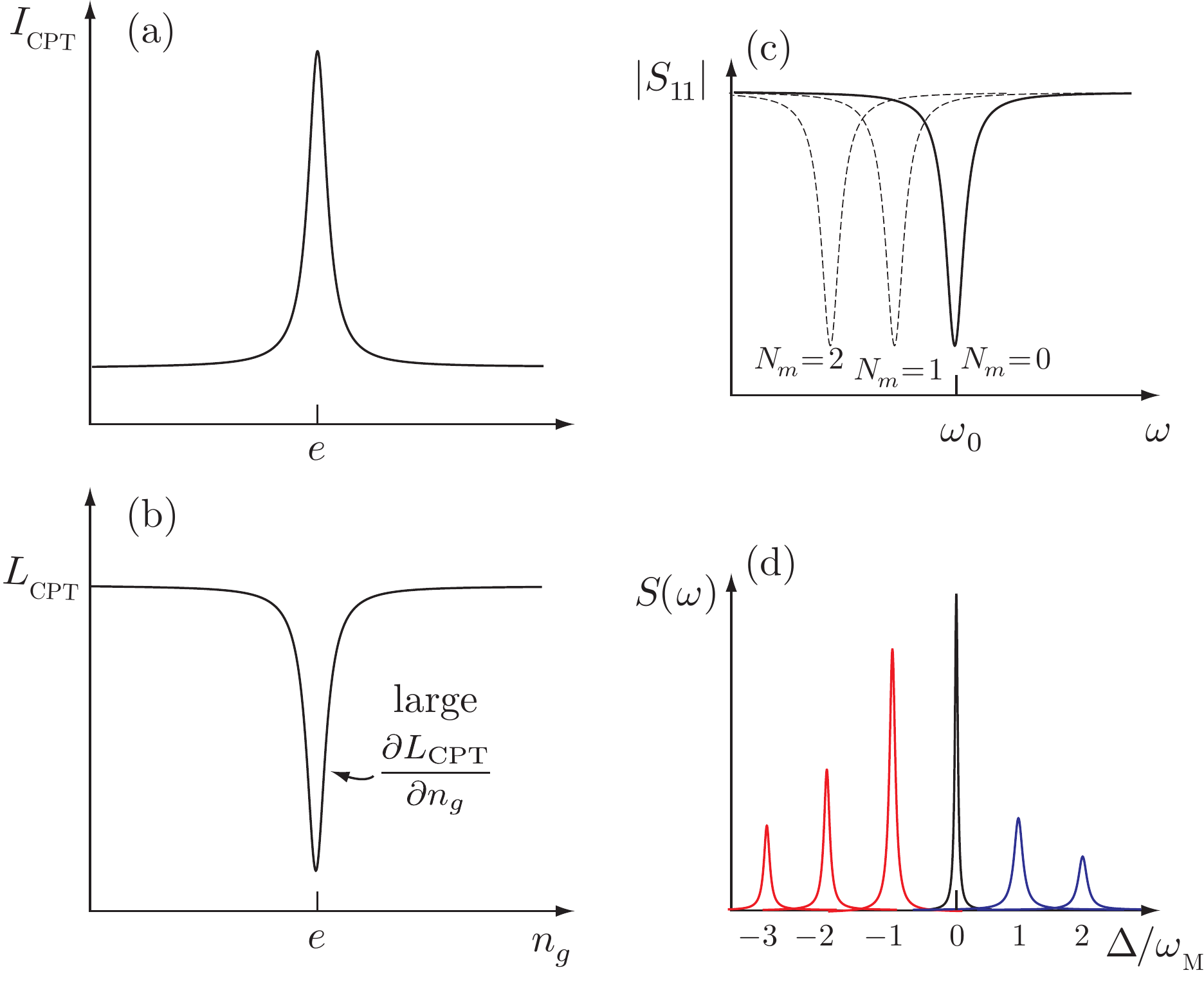}
\caption{\label{scfig} (a)  Schematic illustration of the CPT supercurrent versus gate charge, showing a peak at charge degeneracy. (b) Quantum inductance $L_{\mathrm{CPT}}$ versus gate charge showing a sharp dip at the supercurrent maximum. (c) Potential dispersive shift of cavity resonance for different phonon numbers. (d) Power spectrum of reflected light showing both red and blue detuned resonances associated with mechanical quanta. }
\end{center}
\end{figure}

The above-described cCPT functions as a sensor by capacitively coupling the CPT island to  a system of interest, in our case a mechanical resonator (MR) consisting of a doubly clamped beam (made for example  of SiN and coated with Al~\cite{LaHaye:2004,Naik:2006}) as in figure~\ref{mrfig}(b).  An important property of the CPT is that it acts as a charge-tunable quantum inductor $L_{\mathrm{CPT}}$ when biased on its supercurrent branch; $L_{\mathrm{CPT}}$ is the kinetic inductance associated with the CPT's gate charge dependent supercurrent~\cite{Matveev:1993} [see figure~\ref{scfig}(a) and (b)].  When the CPT is embedded in a microwave cavity, $L_{\mathrm{CPT}}$ appears in parallel with the cavity's effective inductance $L$ at resonance, as in figure~\ref{mrfig}(c), and can therefore cause a dispersive shift of the cavity resonant frequency.  When the CPT island is capacitively coupled to a charged mechanical resonator, motion of the resonator can modulate $L_{\mathrm{CPT}}$ and therefore shift the cavity frequency $\omega_0$.  This dispersive measurement scheme is closely related to that demonstrated in the inductive single electron transistor~\cite{Sillanpaa:2004,Sillanpaa:2005c}.

The key question is how large a shift in $\omega_0$ will result from motion of the MR.  It is straightforward to estimate
\[
\frac{\partial\omega_0}{\partial n_g} =  \frac{\partial \omega_0}{\partial  L_{\mathrm{CPT}}}\frac{\partial L_{\mathrm{CPT}}}{\partial n_g}=-\frac{\omega_{0}}{2}\frac{L}{ L_{\mathrm{CPT}}} g_{\mathrm{CPT}}
\]
where $g_{\mathrm{CPT}}=\frac{1}{ L_{\mathrm{CPT}}}\frac{\partial L_{\mathrm{CPT}}}{\partial n_g }$ is the effective gain of the CPT~\cite{Sillanpaa:2005c} and we have assumed $ L_{\mathrm{CPT}}\gg L$.  Using  realistic numbers for the CPT and cavity ($\omega_0=2\pi\times{5}~{\mathrm{GHz}}$, $L={2}~{\mathrm{nH}}$, $L_{\mathrm{CPT}}={50}~{\mathrm{nH}}$ and  $g_{\mathrm{CPT}}=3$), we estimate that  a frequency sensitivity of
\[
\frac{\partial\omega_0}{\partial  n_g} = 2\pi\times{300}~{\mathrm{MHz}}/{\mathrm{electron}}
\]
should be readily achievable.

To determine the optomechanical coupling strength $g_0$, we must consider a particular mechanical resonator.  Here we envisage using a small doubly clamped beam about ${10}~\mu{\mathrm{m}}$ long  with a mass $m = {0.2}~{\mathrm{pg}}$ and a mechanical  resonant frequency  $\omega_{m} \approx 2 \pi\times{10}~{\mathrm{MHz}}$.  Such resonators are relatively easy to fabricate out of high-stress SiN film~\cite{Verbridge:2006,Verbridge:2008,Hertzberg:2009}, have been successfully coupled to SETs~\cite{LaHaye:2004,Naik:2006}, and possess a relatively large zero point motion; for the dimensions above,  $x_{\mathrm{zp}}={60}~{\mathrm{fm}}$.  The resonator is metallized so that a large applied dc voltage $V_{\mathrm{MR}}$ couples its motion to the CPT gate charge.  If $x$ is the resonator position,
\[
\frac{\partial  n_g}{\partial x} = \frac{V_{\mathrm{MR}}}{e}\frac{\partial C_{m}}{\partial x}.
\]
Here $C_{m}$ is the coupling capacitance between the CPT and the MR;  $\partial C_{m}/\partial x \approx {2}~{\mathrm{aF}}/{\mathrm{nm}}$, and $V_{\mathrm{MR}}\approx{15}~{\mathrm{V}}$ are achievable numbers~\cite{Rocheleau:2010}.   We then estimate a CPT/nanoresonator coupling of
 \[
 \frac{\partial  n_g}{\partial x}  \approx {200}~{\mathrm{electrons}}/{\mathrm{nm}}.
\]
Combining the above, we estimate that for the cCPT-MR system the optomechanical coupling strength is given by
 \begin{equation}
 g_0 = \frac{\partial\omega_0}{\partial  n_g}\frac{\partial  n_g}{\partial x}  x_{\mathrm{zp}} \approx 2\pi\times{4}~{\mathrm{MHz}}.
 \label{goeq}
\end{equation}

For the cCPT, we expect a cavity $Q\approx 10^{4}$, giving $\kappa=2\pi\times{500}~{\mathrm{KHz}}$.  In table~\ref{omtab} we show our resulting estimates for $g_0/\kappa$, $g_0/\omega_m$ and $g_0^{2}/\kappa\omega_m$.  All are of order unity or above, indicating that the cCPT-MR should be well within the single-photon quantum regime. To put these results into context, we also show in table~\ref{omtab}  the same three parameters for similar solid state optomechanical systems~\cite{Teufel:2011a,Chan:2011}, as well as for an atomic system~\cite{Murch:2008}.  Comparing the combined quantum nonlinearity parameter $g_0^{2}/\kappa\omega_m$, we see that the expected value of $3$ for the cCPT-MR {\emph { is roughly seven orders of magnitude greater than that of the nearest solid state systems}}.  Although $g_0^{2}/\kappa\omega_m$ can be even larger in a cold atomic gas~\cite{Murch:2008}, the number of atoms $N$ in such a gas is some five orders of magnitude smaller; in contrast, our focus is on far more macroscopic resonators.

As a first step towards demonstrating that we have entered the single-photon quantum regime, we can measure the power spectrum of light reflected from the cavity when driven at its bare resonance frequency, as in figure~\ref{scfig}(d).  A clear signature of strong coupling would be the appearance of multiple mechanical sidebands in the power spectrum of reflected light, corresponding to absorption or emission of multiple phonons~\cite{Nunnenkamp:2011}.  Note that, in the single-photon ultra-strong coupling regime, it is also possible to read out the cavity photon number using a quantum non-demolition (QND), mechanical displacement measurement scheme~\cite{Jacobs:1994}.  Such a QND measurement approach  necessarily requires \textit{both} $g_{0}/\kappa \gtrsim 1$ and $g_{0}/\omega_{m}\gtrsim 1$, which are satisfied in this cCPT-MR device.

Quantum state tomography on the microwave photons will provide information about the MR state. We can, in particular, employ recently developed tomographic techniques based on quadrature measurements of the cavity output using linear amplifiers~\cite{Eichler:2011,Eichler:2012}.  In combination with quantum state reconstruction~\cite{Lvovsky:2009} using maximum likelihood estimation (MLE) techniques~\cite{Hradil:2004,Hradil:2006}, we expect to be able to reconstruct the density matrix  of the cavity field.  A basic experimental difficulty to  overcome when using phase-preserving linear amplifiers such as the HEMT is that such amplifiers always add noise~\cite{Caves:1982}.  Locating a near-quantum limited superconducting amplifier, e.g., based   on the SLUG (superconducting lumped-element galvanometer--a device closely related to the SQUID)~\cite{Ribeill:2011,Hover:2012}  prior to the HEMT (see figure~\ref{rmfig}), should reduce the number of added noise photons to $\sim 1$.  There should then be significantly less blurring of  the measured quadrature histograms, and comparable improvement in the MLE reconstructions of the cavity photon density matrix  as compared with using just a HEMT alone.  In addition to low noise, the SLUG has a large dynamic range (estimated at up to ${130}~{\mathrm{dB}}$ or more~\cite{Ribeill:2011}), allowing it to accommodate cavity fields containing from only a few to up to a few hundred photons.

\section{Derivation of the Optomechanical Hamiltonian }
\label{Sec:derivation}
In the appendix, we show that the cCPT-MR device can be described by an approximate circuit model with Hamiltonian
\begin{eqnarray}
{\cal{H}}_{\mathrm{cCPT-MR}}&=&\hbar\omega_0 a^{\dagger}{a}+\hbar\omega_m b^{\dagger}{b}+\varepsilon\sigma_z\cr
&&-E_J\sigma_{x} \cos\left[{\Delta}_{\mathrm{zp}} (a + a^{\dagger}) +\pi\Phi_{\mathrm{ext}}/\Phi_{0}\right] + \hbar g_{m} \sigma_{z} (b+b^{\dagger}),
\label{ccptmreq}
\end{eqnarray}
where $\varepsilon=2E_c (1-n_g)$, and where  the CPT-MR coupling is
\begin{equation}
\hbar g_m=\frac{e^2}{C_J}{x_{\mathrm{zp}}}\frac{\partial n_g}{\partial x}=eV_{\mathrm{MR}}\frac{x_{\mathrm{zp}}}{C_J}\frac{\partial C_m}{\partial x}.
\label{gmeq}
\end{equation}
Following the method of reference~\cite{Jacobs:2009}, we group the terms in the Hamiltonian ${\cal{H}}_{\mathrm{cCPT-MR}}$ as follows
\begin{equation}
{\cal{H}}_{\mathrm{cCPT-MR}}=H_0+V+H_{\mathrm{res}},
\label{groupeq}
\end{equation}
where
\begin{equation}
H_0=\varepsilon\sigma_z-E_J\cos\left[\pi\Phi_{\mathrm{ext}}/\Phi_{0}\right]\sigma_{x}
\label{h0eq}
\end{equation}
is the so-called CPT ``auxiliary" system Hamiltonian,
\begin{eqnarray}
V&=&-E_J\left(\cos\left[{\Delta}_0 (a + a^{\dagger}) +\pi\Phi_{\mathrm{ext}}/\Phi_{0}\right]-\cos\left[\pi\Phi_{\mathrm{ext}}/\Phi_{0}\right]\right)\sigma_{x}\cr
&&+ \hbar g_{m} \sigma_{z} (b+b^{\dagger})
\label{perturbeq}
\end{eqnarray}
is viewed as a perturbation to the Hamiltonian $H_0$, and
\begin{equation}
H_{\mathrm{res}}=\hbar\omega_0 a^{\dagger}{a}+\hbar\omega_m b^{\dagger}{b}
\label{reshameq}
\end{equation}
is the resonator Hamiltonian. Since the resonator operator terms $a + a^{\dagger}$ and $b+b^{\dagger}$ appearing in $V$ commute with the auxiliary $H_0$, we can use standard
time-independent perturbation theory to diagonalize $H_{\mathrm{aux}}=H_0+V$ and in particular approximately determine its energy eigenvalues $E_n$. Assuming that the auxiliary system is in its lowest energy eigenstate, with eigenvalue $E_1$,  yields an approximate, ``engineered" Hamiltonian describing the interacting microwave and mechanical resonator resonators: $H_{\mathrm{eng}}=H_{\mathrm{res}}+E_1$. Solving for $E_1$ to second order in $V$, we obtain:
\begin{eqnarray}
&&H_{\mathrm{eng}}=\hbar\omega_0 a^{\dagger}{a}+\hbar\omega_m b^{\dagger}{b}\cr
&&-\frac{E_J^2}{E_0}\cos\left(\pi\Phi_{\mathrm{ext}}/\Phi_{0}\right)\left(\cos\left[{\Delta}_{\mathrm{0}} (a + a^{\dagger}) +\pi\Phi_{\mathrm{ext}}/\Phi_{0}\right]-\cos\left[\pi\Phi_{\mathrm{ext}}/\Phi_{0}\right]\right)\cr
&&-\frac{(\varepsilon E_J)^2}{2 E_0^3}\left(\cos\left[{\Delta}_{\mathrm{0}} (a + a^{\dagger}) +\pi\Phi_{\mathrm{ext}}/\Phi_{0}\right]-\cos\left[\pi\Phi_{\mathrm{ext}}/\Phi_{0}\right]\right)^2\cr
&&- \frac{\varepsilon}{E_0}\hbar g_m (b+b^{\dagger})-\frac{ E_J^2}{2 E_0^3}(\hbar g_m)^2 \cos^2\left(\pi\Phi_{\mathrm{ext}}/\Phi_{0}\right) (b+b^{\dagger})^2 \cr
&&+\frac{\varepsilon E_J^2}{E_0^3}\hbar g_m\cos\left(\pi\Phi_{\mathrm{ext}}/\Phi_{0}\right)\cr
&&\times\left(\cos\left[{\Delta}_{\mathrm{zp}} (a + a^{\dagger}) +\pi\Phi_{\mathrm{ext}}/\Phi_{0}\right]-\cos\left[\pi\Phi_{\mathrm{ext}}/\Phi_{0}\right]\right)(b+b^{\dagger}),
\label{hengeq}
\end{eqnarray}
where $E_0=\sqrt{\varepsilon^2+E_J^2\cos^2\left(\pi\Phi_{\mathrm{ext}}/\Phi_{0}\right)}$, such that $2E_0$ gives the energy level splitting for the unperturbed CPT Hamiltonian  $H_0$.  From equation~(\ref{hengeq}), we see that the CPT effects a gate voltage and flux tunable interaction between the microwave and mechanical oscillators, as well as self-interactions for the two oscillators. The method we have used is expected to provide good approximations to the oscillator interactions provided the CPT level splitting satisfies $E_0\gg \hbar\omega_0,\hbar\omega_m$, so that the dynamics of the CPT is effectively frozen out. The perturbation expansion to second order in $V$ will apply when $\hbar g_m\ll E_J$, which will indeed generally be the case, and when the cavity photon number, $n=\langle a^{\dagger}a\rangle$, is sufficiently small to ensure that $\Delta_0\sqrt{n}\ll 1$.

The Hamiltonian given by equation~(\ref{hengeq}) takes a variety of different forms for different choices of the external flux. The usual optomechanical interaction is recovered if we set $\Phi_{\mathrm{ext}}=0$ and, assuming sufficiently small $n$, we expand the cosine terms in~(\ref{hengeq}) keeping terms up to second order in $\Delta_0(a + a^{\dagger})$ overall. Applying a rotating wave approximation to the terms of the form $(a + a^{\dagger})^2$  (an approach which will be valid provided any external drives applied are close to the cavity frequency), we obtain
\begin{eqnarray}
H_{\mathrm{eng}}&=&\left(\hbar\omega_0-\frac{E_J^2}{E_0}{\Delta}_0^2\right) a^{\dagger}{a}+\hbar\omega_m b^{\dagger}{b}+\frac{\varepsilon E_J^2}{E_0^3}\hbar g_m{\Delta}_0^2a^{\dagger}a(b+b^{\dagger})\nonumber \\
&&- \frac{\varepsilon}{E_0}\hbar g_m (b+b^{\dagger})-\frac{ E_J^2}{2 E_0^3}(\hbar g_m)^2 (b+b^{\dagger})^2,
\label{hengeq2}
\end{eqnarray}
where now $E_0=\sqrt{\varepsilon^2+E_J^2}$.
This Hamiltonian can be simplified further by noting that terms of the form $(b+b^{\dagger})$ simply lead to a displacement of the mechanical resonator whilst the term in $(b+b^{\dagger})^2$ renormalizes its frequency a little. There is also a slight renormalization of the cavity frequency. Thus we finally obtain the standard optomechanical Hamiltonian, equation~(\ref{homeq}). An expansion of the cosine terms that retained terms of order $\Delta_0^4(a + a^{\dagger})^4$ overall would also lead to a Kerr nonlinearity in the cavity~\cite{Haroche:2006}, but of course this would be a small correction in the regime of low photon numbers in which we are working.

The vacuum optomechanical coupling strength is
\begin{equation}
 g_0 =\frac{\varepsilon E_J^2}{E_0^3}\Delta_0^2 g_m.
\label{microg0eq}
\end{equation}
The factor ${\varepsilon E_J^2}/{E_0^3}$ varies with $n_g$ in a way which matches the gradient of $L_{\mathrm{CPT}}$ shown in figure~\ref{scfig}. It reaches a maximum magnitude of $\sqrt{4/27}$ (independent of $E_c$ and $E_J$) when $n_g=1\pm E_J/(2\sqrt{2}E_c)$. Using the parameters in section~\ref{Sec:2}, along with a cavity impedance $Z_{0}=120~\Omega$ and junction capacitance $C_{\rm J}=0.32~\rm{aF}$, we get $g_m=2\pi\times 7~ \rm{MHz}$ and $\Delta_{0}\approx 0.07$. For the optimal choice of $n_g$ one then obtains an ultra-strong coupling $g_0=2\pi\times 3~\rm{MHz}$.  Therefore, the relevant parameters have the values $g_{0}/\kappa\sim 6$, $g_{0}/\omega_{m}\sim0.3$, and $g^{2}_{0}/\kappa\omega_{m}\sim 1.8$, which are consistent with the physical estimates obtained in section~\ref{Sec:2}.

\section{Discussion}
\label{Sec:disc}
While the standard optomechanical Hamiltonian~(\ref{homeq}) can be recovered by approximation from the CPT-engineered, microwave-mechanical oscillator Hamiltonian~(\ref{hengeq}), it is important to note that, by tuning the flux, one can access a broader class of strong optomechanical interactions. In particular, for non-zero $\Phi_{\mathrm{ext}}$ (e.g., $\Phi_{\mathrm{ext}}=\Phi_0/4$), a bilinear interaction term $(a+a^{\dagger})(b+b^{\dagger})$ is also present. Such a broader class of tunable interactions may facilitate the generation and detection of a correspondingly broad class of mechanical resonator quantum states.

One of our main goals in future work is to determine if it is possible to generate {\emph{steady-state}} quantum behavior in the mechanical resonator under ``warm" conditions, i.e., $\hbar\omega_m <k_BT$.  Several recent studies~\cite{Rodrigues:2010,Nunnenkamp:2011,Rips:2012,Nation:2013,Lorch:2013} indicate that steady-state mechanical quantum behavior may well be possible, provided the thermal excitations are minimal.  However,  at the base temperature of a dilution refrigerator, a mesoscale $\sim 10~{\mathrm{MHz}}$  mechanical resonator will be occupied by some one hundred or so phonons on average.   When the mechanical frequency $\omega_m$ is greater than the cavity linewidth $\kappa$ (the resolved sideband regime, for which $\omega_m/\kappa>1$) it is possible to drive the cavity with a red-detuned signal so as to absorb phonons from the resonator~\cite{WilsonRae:2007,Marquardt:2007}.  This technique has been used in both the optical and microwave multi-photon regimes to cool mechanical resonators to their ground state~\cite{Teufel:2011a,Chan:2011}.  A possible first experimental step would be to extend this approach to the single-photon regime~\cite{Nunnenkamp:2012}.

Once ground state cooling is achieved, we can then investigate the question of how to drive the mechanical resonator into a steady quantum state.  Two approaches suggest themselves.  The first is to apply alternating red-detuned cooling pulses and blue detuned driving pulses. After application of a pulse sequence, the cavity photons are monitored so as to read out the mechanical resonator dynamics.  Here, performing state tomography on cavity photons is expected to be of great benefit.  The second approach is to simultaneously apply cooling and driving pulses; such a technique has been proposed for generating steady quantum states in a non-linear mechanical resonator~\cite{Rips:2012}, and has been used for back-action evading measurements in the multiphoton regime~\cite{Hertzberg:2009}.

\section*{Acknowledgments}

While the present manuscript was in preparation, we became aware of reference~\cite{heikkila:2013}, which discusses a very similar scheme to ours. AJR and MPB  were supported by the NSF (grants DMR-1104821 and DMR-1104790) and by AFOSR/DARPA agreement FA8750-12-2-0339. ADA was supported by the EPSRC (UK), Grant No. EP/I017828. PDN was supported by startup funding from Korea University.

\appendix

\section{Derivation of the cCPT-MR Circuit Model}
\begin{figure}
\begin{center}
\includegraphics[width=8.5cm]{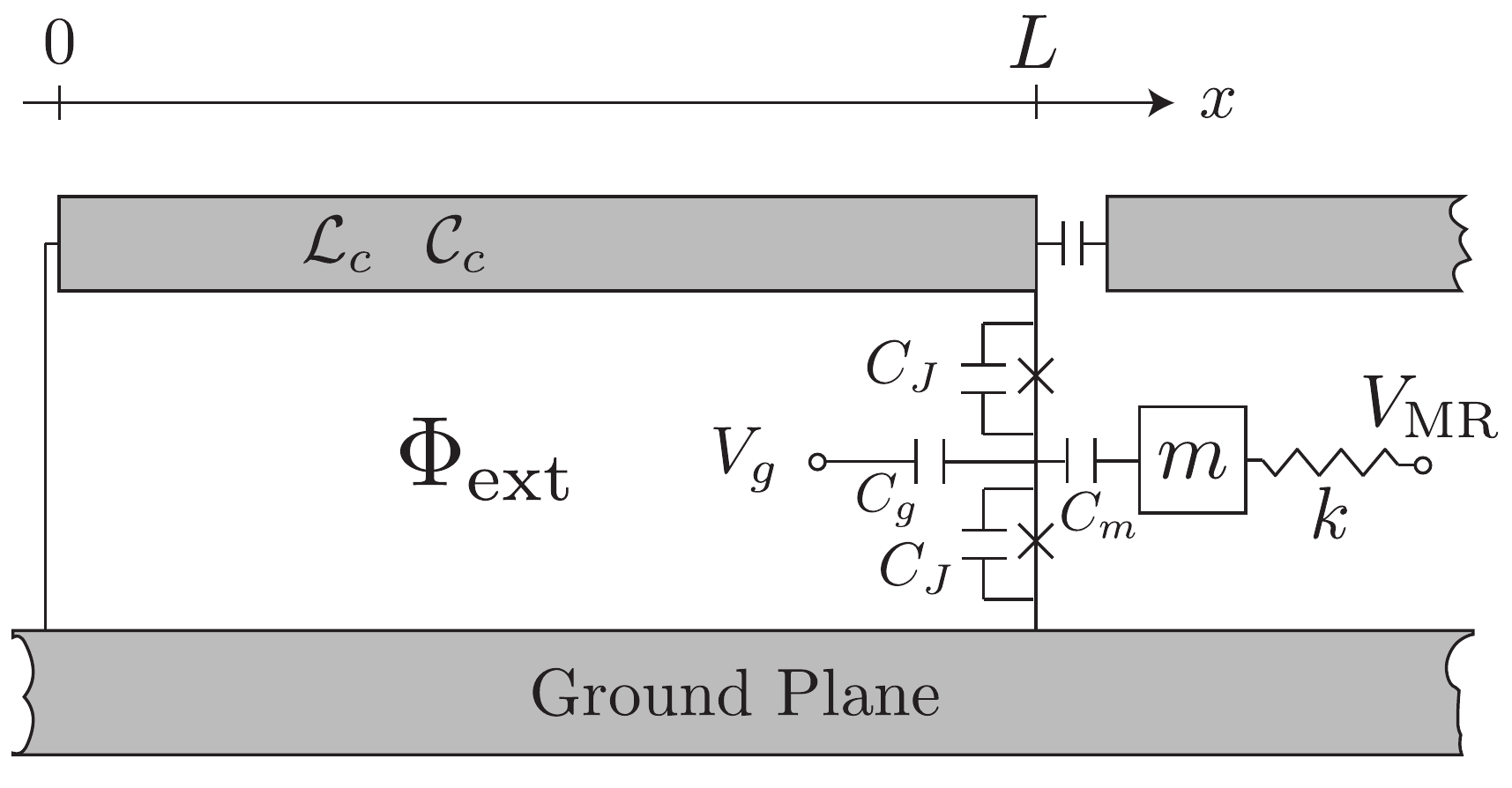}
\caption{\label{cfig} Simplified model of the cCPT-MR  system, where the cavity center conductor has length $L$, and the Josephson junctions are assumed to have equal capacitances $C_J$ and critical currents $I_c$. The cavity inductance and capacitance per unit length are denoted ${\mathcal{L}_c}$, ${\mathcal{C}_c}$, respectively. Note that the   center conductor is shorted to ground at the $x=0$ end and is weakly
coupled via a capacitor to a probe/transmission line at the $x=L$ end.}
\end{center}
\end{figure}

In this appendix, we give a derivation of the circuit model of the cCPT-MR device. Referring to figure~\ref{cfig}, we approximate the microwave cavity for the lowest modes as a one-dimensional strip of length~$L$.
Kirchhoff's laws  yield the following equations in terms of the CPT phases $\gamma_{\pm}(t)=(\varphi_1(t)\pm\varphi_2(t))/2$ (with $\varphi_1$, $\varphi_2$ the gauge invariant phases across the Josephson junctions), and the cavity phase field $\phi_c(x,t)$:
\begin{equation}
2 C_J \frac{\Phi_0}{2\pi}\frac{d^2\gamma_+}{dt^2} +2 I_c \sin\gamma_+\cos\gamma_- -C_g \frac{d V_g}{dt}-C_m \frac{d V_{\mathrm{MR}}}{dt}=-\frac{\Phi_0}{\pi {\mathcal{L}}_c}\left.\frac{\partial \phi_c}{\partial x}\right|_{L} ,
\label{gamma+eq}
\end{equation}
\begin{equation}
2 C_J \frac{\Phi_0}{2\pi}\frac{d^2\gamma_-}{dt^2} +2 I_c \cos\gamma_+\sin\gamma_- +C_g \frac{d V_g}{dt}+C_m \frac{d V_{\mathrm{MR}}}{dt}=0,
\label{gamma-eq}
\end{equation}
\begin{equation}
\frac{\partial^2 \phi_c}{\partial t^2}=({\mathcal{L}}_c{\mathcal{C}}_c)^{-1}\frac{\partial^2\phi_c}{\partial x^2},~0<x<L,
\label{cavitywaveeq}
\end{equation}
where we neglect the coupling to the probe line, since we are concerned here only with deriving the closed system circuit model Hamiltonian~(\ref{ccptmreq}).
The boundary  condition at $x=0$ can be written as
\begin{equation}
\phi_c(0,t)=0,
\label{x=0boundaryeq}
\end{equation}
while the  junction condition at $x=L$ is
\begin{equation}
2\gamma_+(t)-\phi_c(L,t)=2\pi n +2\pi\Phi/\Phi_0,
\label{junctioneq}
\end{equation}
where $n$ is an integer and $\Phi$ is the flux threading the superconducting loop formed out of the center conductor, ground plane, and the CPT. In the following, we will approximate the flux as $\Phi\approx\Phi_{\mathrm{ext}}$, i.e., assume that the induced flux in the loop due to the circulating super current can be neglected. We  also ``freeze" out the MR motion, so that $C_m$ is fixed and non-dynamical; the mechanical component is  straightforwardly introduced once we have obtained the cCPT Hamiltonian~(\ref{heq}).

We  now use equation~(\ref{junctioneq}) to eliminate $\gamma_+$ from the dynamical equations; equations~(\ref{gamma-eq}) and (\ref{gamma+eq}) become respectively
\begin{eqnarray}
2 C_J \frac{\Phi_0}{2\pi}\frac{d^2\gamma_-}{dt^2} &+&2 I_c \cos\left[\phi_c(L,t)/2 +\pi\Phi_{\mathrm{ext}}/\Phi_0\right]\sin\gamma_-\cr
&& +C_g \frac{d V_g}{dt}+C_m \frac{d V_{\mathrm{MR}}}{dt}=0
\label{gamma-2eq}
\end{eqnarray}
and
\begin{eqnarray}
\phi'_c(L,t)+\frac{C_J}{2{\mathcal{C}_c}}\phi''_c(L,t)&=&-\frac{2\pi {\mathcal{L}_c}I_c}{\Phi_0}\sin\left[\phi_c(L,t)/2+\pi\Phi_{\mathrm{ext}}/\Phi_0\right]\cos\gamma_- \cr
&&+\frac{\pi {\mathcal{L}_c}}{\Phi_0} \left(C_g\dot{V}_g+C_m\dot{V}_{\mathrm{MR}}\right),
\label{x=Lboundaryeq}
\end{eqnarray}
where have set $n=0$ since it does not affect the observable dynamics and we have used the cavity wave equation~(\ref{cavitywaveeq}) to replace $\ddot{\phi}_c$ with ${\phi}''_c$.
Equation~(\ref{x=Lboundaryeq}) is interpreted as a (rather nontrivial) boundary condition on the cavity field $\phi_c(x,t)$ at the $x=L$ end that couples the cavity to the CPT.

We now formally solve the cCPT equations~(\ref{cavitywaveeq}) and (\ref{gamma-2eq}), subject to the boundary conditions~(\ref{x=0boundaryeq}) and (\ref{x=Lboundaryeq}), using the  approximate eigenfunction expansion method, with equation~(\ref{x=Lboundaryeq}) replaced by  the following
simpler boundary condition at $x=L$:
\begin{equation}
\phi'_c(L,t)+\frac{C_J}{2{\mathcal{C}_c}}\phi''_c(L,t)\approx\left.\phi'_c(x,t)\right|_{x=L+C_J/(2{\mathcal{C}}_c)}=0,
\label{x=Lboundary2eq}
\end{equation}
expressed approximately as a Neumann boundary condition evaluated at the slightly shifted endpoint $x=L+C_J/(2{\mathcal{C}}_c)$, with $C_J/({\mathcal{C}}_c L)\ll 1$.
We can now apply the method of separation of variables to the cavity wave equation~(\ref{cavitywaveeq}), since the homogeneous boundary conditions~(\ref{x=0boundaryeq}) and  (\ref{x=Lboundary2eq}) define a Sturm-Liouville problem. Neglecting the small endpoint shift  $C_J/{\mathcal{C}}_c\ll L$, the orthogonal eigenfunctions are approximately
\begin{equation}
\phi_n(x)=\sin(k_n x)
\label{eigenfunctioneq}
\end{equation}
and the approximate associated wavenumber eigenvalues are
\begin{equation}
k_n=\frac{\pi (2n+1)}{2L},~n=0,1,2,\dots
\label{eigenvalueeq}
\end{equation}
Note that for the lowest, $n=0$ mode, the mode wavelength is $\lambda/4=L$: hence the name ``$\lambda/4$ resonator".

Proceeding with the eigenfunction expansion method, we assume that solutions $\phi_c(x,t)$ to the wave equation~(\ref{cavitywaveeq}) for $0<x<L$ with the full  boundary conditions~(\ref{x=0boundaryeq}) and (\ref{x=Lboundaryeq}) at $x=0$ and $x=L$, respectively, can be expressed as a series expansion in terms of the eigenfunctions $\phi_n(x)$:
\begin{equation}
\phi_c(x,t)=\sum_n q_n(t)\phi_n(x).
\label{expansioneq}
\end{equation}
From equation~(\ref{expansioneq}) and the orthogonality condition on the $\phi_n$'s, the to be determined time-dependent coefficients $q_n(t)$  are given as
\begin{equation}
q_n(t)=\frac{2}{L}\int_{0}^{L} dx\ \phi_c(x,t) \phi_n(x).
\label{coefficienteq}
\end{equation}
Differentiating~(\ref{coefficienteq}) twice with respect to time and applying the cavity wave equation~(\ref{cavitywaveeq}), we have:
\begin{equation}
\ddot{q}_n(t)=\frac{2}{{\mathcal{L}}_c{\mathcal{C}}_c L}\int_{0}^{L} dx\ \phi''_c(x,t) \phi_n(x).
\label{coefficient2eq}
\end{equation}
Integrating~(\ref{coefficient2eq}) by parts twice, applying the boundary conditions~(\ref{x=0boundaryeq}) and (\ref{x=Lboundaryeq}) on $\phi_c(x,t)$ (with shift term $C_J/{\mathcal{C}}_c$ neglected), the  eigenvalue equation $\phi_n''(x)=-k_n^2 \phi_n(x)$ and also equation~(\ref{coefficienteq}), we obtain
\begin{eqnarray}
\ddot{q}_n(t)&=&-\omega^2_n q_n(t)-\frac{4\pi I_c}{\Phi_0{\mathcal{C}}_c L} \cos\gamma_-\sin\left[\frac{1}{2}\sum_{n'} q_{n'} (t)+\pi\Phi_{\mathrm{ext}}/{\Phi_0}\right] \cr
&&+\frac{2\pi}{\Phi_0{\mathcal{C}}_c L} \left(C_g \dot{V}_g+C_m \dot{V}_{\mathrm{MR}}\right),
\label{coefficient3eq}
\end{eqnarray}
where the free cavity mode oscillator frequencies are
\begin{equation}
\omega_n^2=\frac{k_n^2}{{\mathcal{L}}_c{\mathcal{C}}_c}.
\label{modefrequencyeq}
\end{equation}
In terms of the cavity mode phase coordinates $q_n(t)$, the $\gamma_-$ equation~(\ref{gamma-2eq}) becomes
\begin{eqnarray}
2 C_J \frac{\Phi_0}{2\pi}\frac{d^2\gamma_-}{dt^2} &+&2 I_c \sin\gamma_-\cos\left[\frac{1}{2}\sum_{n} q_{n} (t)+\pi\Phi_{\mathrm{ext}}/{\Phi_0}\right] \cr
&&+C_g \frac{d V_g}{dt}+C_m \frac{d V_{\mathrm{MR}}}{dt}=0.
\label{gamma-3eq}
\end{eqnarray}

The closed cCPT system equations of motion~(\ref{coefficient3eq}) and (\ref{gamma-3eq}) follow from the Hamiltonian
\begin{eqnarray}
H&=&\left(\frac{2\pi}{\Phi_0}\right)^2\sum_n\frac{1}{2 C_n}\left(p_n+\frac{\Phi_0}{4\pi}e n_g\right)^2+\left(\frac{\Phi_0}{2\pi}\right)^2\sum_n\frac{q_n^2}{2L_n} \cr
&& +4 E_c (N-n_g/2)^2 -2 E_J \cos\gamma_-\cos\left[\frac{1}{2}\sum_{n} q_{n} +\pi\Phi_{\mathrm{ext}}/{\Phi_0}\right],
\label{hamiltonianeq}
\end{eqnarray}
where  $N=p_-/\hbar$ is minus the number of excess Cooper pairs on the island, $n_g=(C_g V_g+C_m V_{\mathrm{MR}})/e$ is the polarization charge induced by the applied gate voltage biases $V_g$ and $V_{\mathrm{MR}}$, $E_c=e^2/(2C_J)$ is the approximate CPT charging energy (neglecting $C_g$), and $E_J =I_c\Phi_0/(2\pi)$ is the Josephson energy of a single JJ. The lumped capacitance and inductance elements are defined as  $C_n={\cal{C}}_c L/2$ and $L_n=1/(\omega_n^2 C_n)$, respectively.

Hamiltonian~(\ref{hamiltonianeq}) describes the closed cCPT system, approximate discrete mode  classical dynamics. In modeling the experiment, the various circuit  lumped element parameters appearing in~(\ref{hamiltonianeq}) can be selected so as to provide the best fit to the device characteristics. In this way,  Hamiltonian~(\ref{hamiltonianeq}) is assumed to be more versatile than the original starting equations at the beginning of this section, which are tied to a particular model of the cavity geometry.

In terms of the Cooper pair island number eigenbasis, the quantum Hamiltonian corresponding to equation~(\ref{hamiltonianeq}) can be written as
\begin{eqnarray}
{\mathcal{H}}&=&\sum_n\hbar\omega_n a_n^{\dagger}a_n +4 E_{c}\sum_{N=-\infty}^{+\infty}\left(N-n_g/2\right)^2|N\rangle\langle N|\cr
&&-E_J\sum_{N=-\infty}^{+\infty}\left(|N+1\rangle\langle N|+|N-1\rangle\langle N|\right)\cr
&&\cdot\cos\left[\sum_n\Delta_n (a_n+a_n^{\dagger}) +\pi\Phi_{\mathrm{ext}}/{\Phi_0}\right],\label{quantumhamiltonianeq}
\end{eqnarray}
where we have neglected the gate voltage dependent term in the cavity mode coordinate part of the Hamiltonian and where $\Delta_n$ is the zero-point uncertainty of the cavity mode phase  coordinate $q_n$:
\begin{equation}
\Delta_n=\sqrt{\frac{\pi\sqrt{L_n/C_n}}{R_K}}=\sqrt{\frac{Z_n}{R_K}},
\label{zeropointeq}
\end{equation}
with $Z_n$ the cavity mode impedance and $R_K=h/e^2\approx 25.8~{\mathrm{k}}\Omega$ the von Klitzing constant. Restricting to the lowest, $n=0$ cavity mode and truncating  to a two-dimensional subspace involving linear combinations of only zero ($|0\rangle$) and one ($|1\rangle$) excess Cooper pairs on the island then yields the cCPT Hamiltonian~(\ref{heq}) given in the main text. The cCPT-MR Hamiltonian~(\ref{ccptmreq}) then follows from (\ref{heq}) by inserting the MR free Hamiltonian $\hbar\omega_m b^{\dagger} b$ and Taylor expanding the bias voltage $V_{\mathrm{MR}}$, and in turn $n_g$, to first order in the MR displacement to give the optomechanical coupling.

\section*{References}

\bibliography{Rimberg}

\end{document}